\documentclass[aps,prd,10pt,twocolumn,superscriptaddress]{revtex4}
\usepackage[dvipsnames]{xcolor}
\usepackage{physics}
\usepackage{braket}
\usepackage{scalerel}
\usepackage{blindtext} 
\usepackage{epsfig, cancel}

\usepackage{latexsym}
\usepackage{natbib, comment}
\usepackage{mathrsfs,amsmath,amssymb,amsthm,amsfonts,tikz,graphicx,accents,hyperref,color}
\usepackage{url}
\usepackage{dcolumn}
\usepackage{multirow}
\usepackage{color}
\usepackage{cancel}
\usepackage{soul}
\usepackage[normalem]{ulem}
\usepackage{txfonts}
\usepackage{epsfig}
\usepackage{psfrag}
\usepackage{subfigure}
\hypersetup{colorlinks=true}
\usepackage{mathtools}
\usepackage{enumitem}
\usepackage{float}
\usepackage{caption,ragged2e}
\usepackage[normalem]{ulem}

\def\H0{{\text{H}\hspace*{-2.05mm}\text{H} 0\hspace*{-1.35mm}0\ }}

\usepackage[all]{xy}

\usepackage{slashed}
\usepackage{slashed,ccaption}
\usepackage{multirow}

\usepackage{caption}

\usepackage{array}
\hypersetup{ linktoc=all,
    colorlinks, linkcolor={palatinateblue},
    citecolor={red}, urlcolor={amaranth} 
}

\graphicspath{{Images/}}

\DeclareSymbolFont{extraup}{U}{zavm}{m}{n}
\DeclareMathSymbol{\varheart}{\mathalpha}{extraup}{86}
\DeclareMathSymbol{\vardiamond}{\mathalpha}{extraup}{87}
\makeatletter
\renewcommand*{\@fnsymbol}[1]{\ensuremath{\ifcase#1\or \clubsuit \or \vardiamond \or \varheart\or
    \spadesuit\or \mathparagraph\or \|\or **\or \dagger\dagger
    \or \ddagger\ddagger \else\@ctrerr\fi}}
\makeatother

\definecolor{rosy}{RGB}{230,235,252}
\definecolor{myframetitle}{RGB}{90,89,170}
\definecolor{myblocktitle}{RGB}{140,185,249}
\definecolor{mytitle}{RGB}{10,80,26}

\definecolor{darkgreen}{RGB}{27,130,45}
\definecolor{darkblue}{rgb}{0,0,0.3}
\definecolor{darkred}{rgb}{0.7,0,0}

\definecolor{light gray}{RGB}{220,220,220}
\definecolor{dark purple}{RGB}{108,0,217}
\definecolor{pink}{RGB}{190,20,100}
\definecolor{orang}{RGB}{193,63,0}
\definecolor{green}{RGB}{11,98,17}
\definecolor{darkpink}{RGB}{153,0,76}
\definecolor{bluegreen}{RGB}{0,102,102}
\definecolor{greenlagan}{RGB}{0,102,0}
\definecolor{redgreen}{RGB}{102,102,0}
\definecolor{Redgreen}{RGB}{153,76,0}
\definecolor{vividviolet}{rgb}{0.62, 0.0, 1.0}
\definecolor{amaranth}{rgb}{0.9, 0.17, 0.31}
\definecolor{palatinateblue}{rgb}{0.15, 0.23, 0.89}
\definecolor{brightpink}{rgb}{1.0, 0.0, 0.5}
\definecolor{cornflowerblue}{rgb}{0.39, 0.58, 0.93}
\definecolor{deepcarminepink}{rgb}{0.94, 0.19, 0.22}
\definecolor{radicalred}{rgb}{1.0, 0.21, 0.37}

\newcommand\ignore[1]{}
\usepackage[most]{tcolorbox}

\tcbset{highlight math style={left=02mm,right=02mm,top=02mm,bottom=02mm}} 
\usepackage{empheq}

\newcommand\inbox[1]{\tcbset{fonttitle=\scriptsize} \tcboxmath[colback=white,colframe=black!70]{#1}}

\begin{document}

\newcommand{\mytitle}{\begin{center}{\large{\textbf{Generalized Symmetries in Shallow Water}}}
\end{center}}

\title{{\mytitle}}
\author{V.~Taghiloo}\email{v.taghiloo@iasbs.ac.ir}
\affiliation{School of Physics, Institute for Research in Fundamental
Sciences (IPM), P.O.Box 19395-5531, Tehran, Iran}
\affiliation{Department of Physics, Institute for Advanced Studies in Basic Sciences (IASBS),
P.O. Box 45137-66731, Zanjan, Iran}
\author{M.H.~Vahidinia}\email{ vahidinia@iasbs.ac.ir}
\affiliation{Department of Physics, Institute for Advanced Studies in Basic Sciences (IASBS),
P.O. Box 45137-66731, Zanjan, Iran}
\affiliation{School of Physics, Institute for Research in Fundamental
Sciences (IPM), P.O.Box 19395-5531, Tehran, Iran}

\begin{abstract}
Recent developments have extended the concept of global symmetries in several directions, offering new perspectives across a wide range of physical systems. This work shows that generalized global symmetries naturally emerge in shallow water systems. In particular, we demonstrate that two subsystem symmetries—previously studied primarily in exotic field theories—arise intrinsically in the dynamics of shallow water flows. A central result is that the \textit{local} conservation of \textit{potential vorticity} follows directly from the {first subsystem symmetry}, revealing that the classic Kelvin circulation theorem is rooted in these symmetries. Notably, the associated charge algebra forms a Kac-Moody current algebra, with the level determined by the spatial variation of the Coriolis parameter. {Beyond the first subsystem symmetry, we also identify a second one,} construct the corresponding Noether charges, and explore their potential applications.
\end{abstract}
\maketitle

\section{\label{intro} Introduction}
The ocean and the atmosphere are the two deepest fluid systems on Earth. However, it may seem surprising that they are classified as shallow water systems. The reason is simple: while their depths are immense compared to everyday scales, they are relatively small compared to their vast horizontal extent. This is precisely the defining feature of shallow water systems—the depth is negligible in comparison to the horizontal scale. In these systems, vertical evolution is captured by a single scalar field, the height field, while explicit dependence on the vertical dimension is neglected. This simplification, known as the shallow water approximation, reduces the system’s dimensionality from a fully three-dimensional ($1+3$) framework to an effectively two-dimensional ($1+2$) description, making the analysis significantly more tractable.  

The dynamics of linear shallow water systems are governed by the height field $\eta(t, \vb{x})$ and the horizontal velocity field $u_i(t, \vb{x})$, where $\vb{x} = (x, y)$ denotes the horizontal coordinates and $i \in \{x, y\}$ labels the velocity components. These fields obey the following equations \cite{vallis_2017, Fluid-Mech-Book-1, Tong-lectures} \footnote{Note that throughout this paper we do not distinguish between upper and lower indices, as we work in two-dimensional Cartesian coordinates.}  
\begin{subequations}\label{lin-shallow-eom-f0}
    \begin{align}
        &\partial_{t}\eta + H \nabla \cdot u = 0, \\ 
        &\partial_{t}u_i = f_0 \epsilon_{ij} u^j - g \partial_i \eta,  
    \end{align}
\end{subequations}  
where $f_0$ and $g$ denote the constant Coriolis parameter and gravitational acceleration, respectively. These equations can be combined to derive the continuity equations:  
\begin{subequations}\label{Noether-lin-f0}
    \begin{align}
        &\partial_{t}\eta + H \nabla \cdot u = 0, \\ 
        &\partial_{t}\zeta + f_0 \nabla \cdot u = 0,  
    \end{align}
\end{subequations}  
where $\zeta := \epsilon^{ij}\partial_i u_j$ is the fluid vorticity. Remarkably, from these equations, one can identify an intriguing conserved quantity, the \textit{potential vorticity}:  
\begin{equation}\label{cons-PV-f0}
    \partial_{t}\mathcal{Q}_0(t, \vb{x}) = 0, \qquad \mathcal{Q}_0 := H\zeta - f_0 \eta.  
\end{equation}  
Unlike standard continuity equations, where conserved quantities are defined by integration over a spatial slice, $\mathcal{Q}_0$ is conserved \textit{pointwise in space}, representing a genuinely local conservation law.  

Beyond the conservation law Eq. \eqref{cons-PV-f0}, the linearized shallow water Eq. \eqref{lin-shallow-eom-f0} exhibit a notable symmetry: they remain invariant under the transformations  
\begin{subequations}\label{symmetry-tranformation}  
    \begin{align}  
        &\eta(t,\vb{x})\,\, \rightarrow \,\, \eta'(t,\vb{x}) = \eta(t,\vb{x}) - \frac{f_0}{g} c(\vb{x}), \\  
        &u_i(t,\vb{x}) \,\, \rightarrow \,\, u'_i(t,\vb{x}) = u_i(t,\vb{x}) + \epsilon_{ij} \partial_j c(\vb{x}),  
    \end{align}  
\end{subequations}  
where $c(\vb{x})$ is an \textit{arbitrary time-independent} function. These transformations leave the shallow water Eq. \eqref{lin-shallow-eom-f0} unchanged, highlighting an inherent symmetry of the system. The presence of an arbitrary function in the transformation is particularly intriguing and warrants further exploration.

With this brief introduction, we are now in a position to clearly state the aim of this paper. We seek to address the following questions:
\begin{enumerate}  [label=\Roman*.]
    \item Within the framework of Noether's theorem, is there an underlying symmetry that gives rise to the local conservation law Eq. \eqref{cons-PV-f0}?  
    \item Does the symmetry transformation Eq. \eqref{symmetry-tranformation} correspond to any conserved quantities?  
\end{enumerate}

To emphasize once again, the conservation Eq. \eqref{cons-PV-f0} and the symmetry transformations Eq. \eqref{symmetry-tranformation} have distinct characteristics: Eq. \eqref{cons-PV-f0} represents a local conservation law, while the symmetry transformation Eq. \eqref{symmetry-tranformation} involves an arbitrary time-independent function. As we will show, both the underlying symmetry of the continuity Eq. \eqref{cons-PV-f0} and the conserved quantity associated with the symmetry transformation Eq. \eqref{symmetry-tranformation} can be understood within the framework of {\textit{subsystem symmetries} \cite{Paramekanti:2002iup, Vijay:2016phm, Seiberg:2019vrp, Qi:2020jrf}}. For a comprehensive treatment, including reviews, lecture notes, and talks, see \cite{Luo:2023ive, Shao2021, Cordova:2022ruw, Gomes:2023ahz, Schafer-Nameki:2023jdn, Brennan:2023mmt, Bhardwaj:2023kri, Iqbal:2024pee, Shao:2023gho, McGreevy:2022oyu}.

\emph{Outline of the paper.} In Sec. \ref{sec:shallow-water-review}, we review the fundamentals of shallow water systems. Sec. \ref{sec:action} presents an action for the linearized shallow water equations and examines its symplectic structure. In Sec. \ref{sec:Noether}, we explore the symmetries of the shallow water action and derive the corresponding Noether charge within the framework of Noether's theorem. Sec.\ref{sec:applications} discusses the implications of these symmetries. Finally, Sec.\ref{sec:Discussion} provides a summary of our results and offers a discussion of their broader implications.
\section{Shallow Water, Review}\label{sec:shallow-water-review}
In this section, we review both the nonlinear and linear shallow water equations, considering the Coriolis parameter as a general function of spatial coordinates rather than a constant.
\subsection{Shallow water equations}
Shallow water systems are characterized by the height field $\mathrm{H} = \mathrm{H}(t, \vb{x})$ and the horizontal velocity fields $u^i = u^i(t, \vb{x})$, where $\vb{x} = (x, y)$ represents the horizontal coordinates of the fluid. The non-linear dynamical equations governing these systems are given by \cite{vallis_2017, Fluid-Mech-Book-1, Tong-lectures}
\begin{subequations}\label{Non-Shallow-eom}
    \begin{align}
    \frac{D {\mathrm{H}}}{D t}:= \frac{\partial {\mathrm{H}}}{\partial t}+u\cdot \nabla {\mathrm{H}}=&-{\mathrm{H}} \nabla\cdot u\, ,\label{mass-cons}\\
    \frac{D u_i}{D t}:=\frac{\partial u_{i}}{\partial t}+(u \cdot \nabla)u_i=& f(\vb{x}) \epsilon_{ij}u^j-g\, \partial_{i}{\mathrm{H}}\, ,\label{NS-eq}
    \end{align}
\end{subequations}
where, $\frac{D}{Dt}$ represents the material time derivative. These equations have clear physical interpretations:  Eq. \eqref{mass-cons}, represents the conservation of mass, while  Eq. \eqref{NS-eq}, describes Newton's second law, $F = ma$, in the context of fluid dynamics. These equations involve two parameters: $g$, the gravitational acceleration, and $f(\vb{x})$, the Coriolis parameter, which may depend on spatial coordinates.

The shallow water Eq. \eqref{Non-Shallow-eom} are nonlinear differential equations due to the advective term in the material derivatives. In many physical situations, the variations in $\mathrm{H}$ and $u^i$ are small, allowing for a linearization of the equations. This can be achieved by expanding the shallow water Eq. \eqref{Non-Shallow-eom} as follows:
\begin{equation}
    {\mathrm{H}}(t, \vb{x})=H+\eta(t, \vb{x}), \quad u^i(t, \vb{x})=0+u^i(t, \vb{x}),
\end{equation}
where $H$ is a constant height and $\eta \ll H$. The linearized shallow water equations are obtained by retaining only the linear terms in $\eta$ and $u^i$
\begin{subequations}\label{lin-shallow-eom}
    \begin{align}
        &\partial_{t}\eta+H \nabla\cdot u=0 \label{mass-cons-lin}\, ,\\
        &\partial_{t}u_i=f(\vb{x}) \epsilon_{ij} u^j-g\partial_{i}\eta \label{NS-lin}\, .
    \end{align}
\end{subequations}
These are the linearized shallow water equations, introduced in Eq. \eqref{lin-shallow-eom-f0}, generalized from a constant Coriolis parameter $f_0$ to a spatially varying $f(\vb{x})$.

Before proceeding, we briefly discuss the validity of linearizing the shallow water equations. The key criterion is the \textit{Rossby number}:
\begin{equation}
Ro := \frac{U}{f_0 L},
\end{equation}
where $U$ and $L$ are the characteristic velocity and length scales. When $Ro \ll 1$, nonlinear effects are negligible, allowing the use of the linearized equations.

For Earth, with $f_0 \approx 10^{-4} \, \text{s}^{-1}$ and $U \sim 10 \, \text{m/s}$, the Rossby number is $Ro \approx 10^5 \frac{m}{L}$, meaning nonlinear terms are negligible for long-wavelength perturbations. For $L \sim 10^3 \, \text{km}$, $Ro \approx 0.1$.

\emph{Continuity Equations.} It is straightforward to see that Eq. \eqref{lin-shallow-eom} lead to the following two conservation equations
\begin{subequations}\label{Noether-lin}
    \begin{align}
        &\partial_{t}\eta+H\nabla \cdot u=0\, ,\label{h-cons-lin}\\
        & \partial_{t}\zeta+\nabla\cdot (f(\vb{x}) u)=0\, ,\label{vort-cons-lin}
    \end{align}
\end{subequations}
where $\zeta = \epsilon_{ij} \partial_i u_j$ represents the vorticity. Combining these continuity equations leads to an expression describing the time evolution of the potential vorticity $\mathcal{Q}$
\begin{equation}\label{local-pot-vort-cons}
\partial_{t}\mathcal{Q}+H u\cdot\nabla f=0\, , \qquad  \mathcal{Q}:= H\zeta-f \eta\, .
\end{equation}
When the Coriolis parameter is constant, $f = f_0$, the potential vorticity $\mathcal{Q}_0$ satisfies a \textit{local} conservation law, $\partial_t \mathcal{Q}_0 = 0$. As highlighted in the introduction, this result is particularly intriguing due to its inherently local nature \footnote{For similar conservation laws arising from spacetime subsystem symmetries, see \cite{Baig:2023yaz}. For connections to Carrollian symmetries, see \cite{Kasikci:2023tvs}.}. The following sections show that this local conservation equation arises from an underlying subsystem symmetry. For the sake of simplicity, we set $H=1$ in the rest of this paper.

\section{Action for linearized shallow Water}\label{sec:action}
In this section, we introduce an action for the linearized shallow water equations with a position-dependent Coriolis parameter $f(\vb{x})$ Eq. \eqref{lin-shallow-eom}. We begin with the mass conservation Eq. \eqref{mass-cons-lin} and proceed to solve it as follows
\begin{equation}\label{dictionary}
    \eta=\epsilon^{ij}\, \partial_{i} A_{j}\, , \qquad u_{i}= -\ignore{\frac{1}{H}}\, \epsilon_{ij}\, \partial_{t}A_{j}\,.
\end{equation}
As the above equation shows, the Eulerian velocity field $u_i$ is related to the time derivative of $A_i$. This relationship mirrors the connection between velocity and position in elementary physics. The only remaining equation is given by Eq. \eqref{NS-lin}, which, when expressed in terms of $A_i$, takes the following form
\begin{equation}\label{eom-A}
\ignore{H}\,\mathcal{E}_i=\partial_{t}^2 {A}_{i}-f(x)\, \epsilon_{ij}\, \partial_{t}A_{j}-v^{2}[\nabla^{2}A_{i}-\partial_{i}(\nabla \cdot A)]=0\, ,
\end{equation}
where {$v:=\sqrt{g }$, $(H=1)$} is the velocity of {gravity waves}.
We will now introduce an action that reproduces the above equation \cite{Tong:2022gpg}
\begin{equation}\label{action-f}
    \begin{split}
        S[A_i]=\frac{1 \ignore{H^-1}}{2} \int d{}t\, d{}^2 x\, \big[&(\partial_{t}A_{i})^2-v^2\, (\partial_{i}A_j\partial_{i}A_j-\partial_{i}A_j \partial_{j}A_i)\\
        &+f(\vb{x})\epsilon^{ij}A_{i}\, \partial_{t}A_{j}\big]\, .
    \end{split}
\end{equation}
It can be verified that the variation of the above action yields Eq. \eqref{eom-A}.

\paragraph*{{Counting degrees of freedom.}}
{It is instructive to compare the degrees of freedom in the fluid equations of motion \eqref{lin-shallow-eom} with those in the proposed action \eqref{action-f}. The shallow water equations \eqref{lin-shallow-eom} consist of three equations for three variables: the density $\rho$ and the velocity components $u_i$. In contrast, the action \eqref{action-f} is expressed in terms of the field $A_i$, which has only two components. This apparent reduction does not indicate a mismatch, since the mass continuity equation is already solved in formulating the action through the introduction of $A_i$ as in \eqref{dictionary}. As a result, the remaining dynamics are captured by \eqref{NS-lin}, or equivalently by \eqref{eom-A}, which constitute two equations for two unknowns. Therefore, the on-shell physics described by the fluid equations \eqref{lin-shallow-eom} and by the action principle \eqref{action-f} is fully equivalent.}

\paragraph*{{Comparison with \cite{Tong:2022gpg}.}}
A closely related form of the action also appears in Tong’s gauge theory formulation of shallow water equations \cite{Tong:2022gpg}. However, there are important differences between our construction and that of \cite{Tong:2022gpg}. The first difference is that Tong’s action is written for a constant Coriolis parameter $f_{0}$, whereas we extend it to allow for an arbitrary spatial dependence of the Coriolis parameter $f(\vb{x})$. The second, and more significant, difference is that Tong promotes the action \eqref{action-f} (for constant $f_0$) to a gauge theory, namely the Maxwell–Chern–Simons (MCS) action. This comes at a cost: in addition to the fluid equations of motion, one is left with an extra constraint. In Tong’s formulation, this constraint restricts the theory to a narrow subset of solutions, specifically the Poincaré waves. By contrast, in our approach, we do not promote \eqref{action-f} to a gauge theory. This has two advantages: (i) we can accommodate a general Coriolis parameter, and (ii) we avoid the constraint, thereby retaining all linearized solutions of the shallow water system. In fact, for a constant $f_0$, the two actions are related as
\begin{equation}\label{relation-to-Tong}
    S_{\text{MCS}}[A_0, A_i]\big|_{A_0=0} \;=\; S_{\eqref{action-f}}[A_i] \, .
\end{equation}
It is important to emphasize that this relation is not a gauge-fixing procedure. Setting $A_0 = 0$ directly at the level of the action simultaneously eliminates the constraint, and the resulting theory, given by \eqref{action-f}, is no longer a gauge theory.
\paragraph*{Symplectic structure.}
Now, we explore the symplectic structure of the shallow water theory based on the action Eq. \eqref{action-f}. We begin by calculating the first variation of the action
\begin{equation}
    \delta S=\int d{}t d{}^2 x\, \big(-\mathcal{E}_{i}\, \delta A_{i} + \partial_{t}\theta^{t}+\partial_{i} \theta^{i} \big)\, ,
\end{equation}
where $\text{E}_i$ is given by Eq. \eqref{eom-A} and the symplectic potential is as follows
\begin{equation}
    	\ignore{H} \theta^{\, t}=\partial_{t}A_i\, \delta A_i+\frac{1}{2}f\, \epsilon^{ij}\,A_{i}\, \delta A_{j}\, , \qquad 	\ignore{H} \theta^{\, i}=-v^2\, \epsilon^{ij}\, \eta\, \delta A_{j}\, .
\end{equation}
Let us examine the symplectic form on a (partial) Cauchy slice $\Sigma$
\begin{equation}\label{symp-form}
    \Omega=\int_{\Sigma}d^{2}x\, \delta \theta^t=\int_{\Sigma} d^{2}x\, \delta A_i \wedge \delta \pi_i\, ,
\end{equation}
where the canonical momentum conjugate to $A_i$ is given by
\begin{equation}
   	\ignore{H} \pi_i=\partial_{t}A_i-\frac{f}{2}\epsilon_{ij}A_j=	\ignore{H} \epsilon_{ij}u_{j}-\frac{f}{2}\epsilon_{ij}A_j\, .
\end{equation}
From the symplectic form Eq. \eqref{symp-form}, the corresponding equal-time Poisson brackets can be readily extracted
\begin{equation}\label{Poisson-bracket}
    \begin{split}
        &\{A_{i}(t,x),\pi_j(t,x')\}=\delta_{ij}\, \delta^{2}(x-x')\, , \\
        &\{A_{i}(t,x),A_j(t,x')\}=0=\{\pi_{i}(t,x),\pi_j(t,x')\}\, .
    \end{split}
\end{equation}
For the latter convenience, we also introduce the Hamiltonian density associated with the action Eq. \eqref{action-f} 
\begin{equation}\label{Hamiltonian-density}
    \begin{split}
        \mathcal{H}[A_i, \pi_i] &= \frac{1 \ignore{H^-1}}{2}\left((\partial_{t}A_i)^2+v^2 \eta^2\right)=\ignore{\frac{1}{H}}{}\frac{1}{2}\left(	\ignore{H^2}\,u_i^2+v^2 \eta^2\right)\\
        &=\frac{1 \ignore{H^-1}}{2}\left[\left(\pi_i+\frac{f}{2}\epsilon_{ij}A_j\right)^2+v^2\, \eta^2\right]\, .
    \end{split}
\end{equation}

\section{Symmetries of shallow water}\label{sec:Noether}
In this section, we examine the symmetries of the shallow water theory and demonstrate the various types of symmetries of this theory. 
\subsection{{Subsystem symmetry I}}
Consider the following internal subsystem symmetry
\begin{equation}\label{ST-1}
    A_{i}(t,\vb{x}) \to A'_{i}(t,\vb{x}) = A_i(t,\vb{x}) + \partial_{i} \lambda(\vb{x}),
\end{equation}
where $\lambda(\vb{x})$ is an arbitrary \textit{time-independent} function. Under these transformations, the action Eq. \eqref{action-f} changes as follows
\begin{equation}\label{delta-lambda-S}
    \delta_{\lambda}S = \frac{1 \ignore{H^-1}}{2} \int dt\, d^2 x\, \partial_t \left(f(\vb{x})\, \epsilon^{ij}\, \partial_i \lambda\, A_j\right).
\end{equation}
This result clearly shows that the transformation Eq. \eqref{ST-1} is a symmetry of the action Eq. \eqref{action-f}, which we refer to as {\textit{subsystem symmetry I}}. 
{The reason is that, since $\lambda(\vb{x})$ is time-independent and only affects the spatial dependence, it is naturally classified as a subsystem symmetry within the framework of generalized symmetries \cite{Cordova:2022ruw} (see also \cite{Paramekanti:2002iup, Vijay:2016phm, Seiberg:2019vrp, Qi:2020jrf} for the original references).}

{As mentioned in the previous section, the theory \eqref{action-f} is not a gauge theory and does not involve any constraints. This can be further understood in terms of the subsystem symmetry \eqref{ST-1}: since $\lambda(\vb{x})$ is time-independent, it cannot be used to eliminate any components of $A_i(t,\vb{x})$. This makes it clear that $A_i(t,\vb{x})$ cannot be interpreted as a gauge field.}

\paragraph*{{Comparison with \cite{Tong:2022gpg}.}}
Recall \eqref{relation-to-Tong}; the origin of this symmetry can be traced back to the gauge symmetry of the Maxwell–Chern–Simons theory. However, we emphasize that the symmetry transformation \eqref{ST-1} is \textit{not} a gauge symmetry, and, as noted, the action \eqref{action-f} does not define a gauge theory. There is a universal feature at play here: in a gauge theory, if one sets non-dynamical degrees of freedom to zero at the level of the action (for instance, $A_0 = 0$ in Maxwell–Chern–Simons), one loses the associated constraint and thus the corresponding gauge symmetry. The intriguing point, however, is that not all symmetries are lost — a measure-zero subset survives and manifests as a global or physical symmetry in the resulting non-gauge theory. The transformation \eqref{ST-1} can be understood in precisely this way.
\paragraph*{Noether current.} 
We are now ready to compute the Noether current associated with the symmetry transformation Eq. \eqref{ST-1}
\begin{subequations}
    \begin{align}
        &	\ignore{H}\rho_{\lambda}
        =\partial_{t}A_i\, \partial_{i} \lambda+f\, \epsilon^{ij}\,A_{i}\, \partial_{j}\lambda\, ,\\
        &\ignore{H} J^{i}_{\lambda}
        =-v^2\, \eta\, \epsilon^{ij}\, \partial_{j}\lambda\, ,
    \end{align}
\end{subequations}
where we have used $\delta_{\lambda}A_{i}=\partial_{i}\lambda$. It is straightforward to verify that this Noether current is conserved on-shell ($\mathcal{E}_{i}=0$)
\begin{equation}
\partial_{t}\rho_{\lambda}+\partial_{i}J^{i}_{\lambda} = \mathcal{E}_{i}\, \partial_{i}\lambda = 0\, .
\end{equation}
\paragraph*{Noether charge.} The corresponding Noether charge is given by
\begin{equation}
    Q_{\lambda}=\int_{\Sigma} d^{2}x\, \rho_{\lambda}=\ignore{\frac{1}{H}}\int_{\Sigma} d^{2}x\, (\partial_{t}A_i-f\, \epsilon^{ij}\,A_{j})\partial_{i}\lambda\, .
\end{equation}
After performing integration by parts, this quantity, expressed in terms of the fluid variables, is written as follows
\begin{equation}
    \begin{split}
        Q_{\lambda} = & -\ignore{\frac{1}{H}}\int_{\Sigma} d^{2}x\, (\mathcal{Q}-\epsilon^{ij}\, \partial_{i}f\, A_j)\lambda(\vb{x})\\
        &+\ignore{\frac{1}{H}}\int_{\partial \Sigma} dx_i\, \epsilon^{ij}(	\ignore{H} u_j-f A_{j})\lambda(\vb{x})\, ,
    \end{split}
\end{equation}
where $\mathcal{Q}$ in the first line is the potential vorticity given by Eq. \eqref{local-pot-vort-cons}. When the boundary integral vanishes (e.g. when $\partial \Sigma=\varnothing$), we obtain
\begin{equation}\label{Q-lam}
    Q_{\lambda}=-\ignore{\frac{1}{H}}\int_{\Sigma} d^{2}x\, (\mathcal{Q}-\epsilon^{ij}\, \partial_{i}f\, A_j)\lambda(\vb{x})\, .
\end{equation}
It is noteworthy that since $\lambda(\vb{x})$ is an arbitrary function, the integrand itself must also be conserved. This observation motivates the definition of the \textit{generalized potential vorticity charge aspect} as  
\begin{equation}
            \inbox{
            q(t,\vb{x}) = -\ignore{\frac{1}{H}}(\mathcal{Q} - \epsilon^{ij} \partial_i f \, A_j) \, .
            }
\end{equation}
It is straightforward to examine the time evolution to confirm its local conservation
\begin{equation}
\partial_t q = -\ignore{\frac{1}{H}}(\partial_t \mathcal{Q} + 	\ignore{H} \, u \cdot \nabla f )= 0,
\end{equation}
as expected. This result matches Eq. \eqref{local-pot-vort-cons}. When the Coriolis parameter is constant, the charge associated with the {subsystem symmetry I} Eq. \eqref{ST-1} simplifies to the potential vorticity, $q \sim \mathcal{Q}_0$. Finally, we note that the pointwise conservation of potential vorticity can be interpreted as an infinite number of conserved quantities, one for each spatial point, analogous to the charges associated with large gauge transformations in gauge theories.

Thus, we identify the symmetry underlying the local conservation of potential vorticity, addressing the first question (item I) mentioned in the introduction. In summary, the {subsystem symmetry I} Eq. \eqref{ST-1} ensures the local conservation of potential vorticity, as described by Eq. \eqref{local-pot-vort-cons}.

\paragraph*{Charge algebra.} Now we are in a position to compute the charge algebra. To do so, we note that under transformation Eq. \eqref{ST-1}, the potential vorticity charge aspect transforms as 
\begin{equation}\label{q-trans}
    \delta_{\lambda} q=	\ignore{\frac{1}{H}}\epsilon^{ij}\, \partial_{i}f\, \partial_{j}\lambda\, .
\end{equation}
To compute the charge algebra, we use the standard definition of Poisson brackets
\begin{equation}
    \{Q_{\lambda_1}, Q_{\lambda_2}\}=\delta_{\lambda_2}Q_{\lambda_1}\, .
\end{equation}
This formula, together with the transformation law Eq. \eqref{q-trans}, gives
\begin{equation}\label{q-algebra}
            \inbox{
            \{q({t,\vb{x}}), q({t,\vb{x}'})\}=\epsilon^{ij}\, \partial_{i}f(\vb{x})\, \partial_{j}\, \delta^{2}(\vb{x}-\vb{x}')\, .
            }
\end{equation}
Interestingly, the charge algebra turns out to be non-Abelian. Notably, when the Coriolis parameter is constant ($\partial_i f = 0$), the algebra simplifies to a form that is Abelian. Furthermore, the algebra Eq. \eqref{q-algebra} can also be directly derived from the Poisson brackets Eq. \eqref{Poisson-bracket}.
\subsection{{Subsystem symmetry II}}
In this subsection, we explore another class of shallow water symmetries, assuming the Coriolis parameter is constant, $f = f_0$. In this case, the action Eq. \eqref{action-f} exhibits the following global subsystem symmetry 
\begin{equation}\label{ST-2}
    A_{i}(t,\vb{x}) \,\,\, \to \,\,\, A'_{i}(t,\vb{x}) = A_{i}(t,\vb{x}) + \frac{v^2}{f_0} t \, \epsilon^{kl} \partial_{i} \partial_{k} \chi_{l}(\vb{x}) + \chi_{i}(\vb{x}),
\end{equation}
where $\chi_{i}(\vb{x})$ is an arbitrary time-independent vector. Under this transformation, we have the following changes to the observables
\begin{equation}
\delta_{\chi} \eta = \epsilon^{ij} \partial_{i} \chi_{j}\, , \quad \delta_{\chi} u_{i} = -\frac{g}{f_0} \partial_{j} (\partial_{i} \chi_{j} - \partial_{j} \chi_{i})\, .
\end{equation}
{Note that in contrast to the symmetry transformation Eq. \eqref{ST-1}, this transformation modifies observables such as $\eta$ and $u_i$.}

We can decompose $\chi_{i}(\vb{x})$ to the electric and magnetic parts
\begin{equation}\label{decom-A}
\chi_{i}(\vb{x})=\partial_{i}\,\chi_{\tiny_{\mathsf{E}}}+\epsilon_{ij}\,\partial_{j}\,\chi_{\tiny_{\mathsf{M}}}\, .
\end{equation}
In terms of these electric and magnetic functions, the transformations of our dynamical fields are given by
\begin{subequations}
    \begin{align}
        & \delta_{\chi_{\tiny_{\mathsf{E}}}}A_{i}= \partial_{i}\, \chi_{\tiny_{\mathsf{E}}}\,\label{ST-2-M} , \qquad \delta_{\chi_{\tiny_{\mathsf{M}}}}A_{i}=-\frac{v^2\, t}{f_0} \partial_{i}\nabla^2\chi_{\tiny_{\mathsf{M}}}+\epsilon_{ij} \,\partial_{j}\,\chi_{\tiny_{\mathsf{M}}}\, ,\\
        & \delta_{\chi}\eta=-\nabla^2 \chi_{\tiny_{\mathsf{M}}}\, , \qquad \delta_{\chi} u_i=\frac{g}{f_0} \epsilon_{ij}\, \partial_{j} \nabla^2 \chi_{\tiny_{\mathsf{M}}}\, . \label{magnetic-transformation}
    \end{align}
\end{subequations}
The first equation clearly illustrates that the electric part corresponds directly to the {subsystem symmetry I} Eq. \eqref{ST-1}. To isolate the independent charge associated with Eq. \eqref{ST-2}, we focus solely on the magnetic component of Eq. \eqref{decom-A}, which we denote as $\chi$ for clarity (instead of $\chi_{\tiny_{\mathsf{M}}}$), {and refer to it as subsystem symmetry II}. 

The key point to note is that the magnetic transformation corresponds to the symmetry transformation given in Eq. \eqref{symmetry-tranformation}, with \footnote{The symmetry parameter $c(\vb{x})$ has dimensions $[c] = L^2 T^{-1}$ and can be rendered dimensionless using appropriate combinations of $f_0 $, $g$, and $H$.}
\begin{equation}\label{c-chi-rel}
    c(\vb{x})=\frac{g}{f_0} \nabla^{2}\chi(\vb{x})\, .
\end{equation}
To demonstrate that the transformation law Eq. \eqref{ST-2} is indeed a symmetry of the theory described by Eq. \eqref{action-f}, we evaluate the variation of the action under the transformation Eq. \eqref{ST-2}
\begin{gather}
        \hspace{-.1cm}\delta_{\chi}S=\ignore{\frac{1}{H}} \int d{}t d{}^2 x\, \Biggl\{\partial_{t}\left[\frac{f_0}{2}  A_{i}\partial_{i}\chi-v^2\left(\frac{A_i}{f_0}+\frac{t}{2}\epsilon_{ij}A_j\right)\partial_{i}\nabla^2 \chi\right]\nonumber \\
        + v^2\partial_{i}\left(\epsilon_{ij}A_j\, \nabla^2 \chi\right)\Biggr\}.
\end{gather} 
This result conclusively establishes that the transformation Eq. \eqref{ST-2} is a symmetry of the action, since $\delta_{\chi}S$ corresponds to a boundary term. Next, we turn our attention to the Noether charge associated with this transformation.
\paragraph*{Noether current.} The Noether current associated with this symmetry is given by
\begin{subequations}
    \begin{align}
        	\ignore{H}\rho_{\chi}=&
        -f_0\, A_{i}\, \partial_{i}\chi+\epsilon_{ij}\, \partial_{t}A_i\, \partial_{j}\chi\nonumber\\
        &+v^2\left(\frac{1}{f_0}A_i-\frac{t}{f_0}\partial_{t}A_i+t\,  \epsilon_{ij}A_j\right)\partial_{i}\nabla^{2}\chi\, ,\\
        	\ignore{H} J^{i}_{\chi}=&v^2\, \eta\, \left(\frac{v^2  t}{f_0}\epsilon_{ij} \, \partial_{j}\nabla^2 \chi +\partial_{i}\chi\right)-v^2 \, \epsilon_{ij}\, A_{j}\, \nabla^2 \chi\, .
    \end{align}
\end{subequations}
One can check that this current is conserved on-shell ($\mathcal{E}_{i}=0$)
\begin{equation}
    \partial_{t}\rho_{\chi}+\partial_{i}J^{i}_{\chi}\ignore{=-\frac{v^2 t}{f_0} \mathcal{E}_i\, \partial_{i} \nabla^2 \chi+\epsilon_{ij} \mathcal{E}_i\, \partial_{j}\chi\,}=0 .
\end{equation}
\paragraph*{Noether charge.} The corresponding Noether charge is given by
\begin{equation}
    \begin{split}
        Q_{\chi} =\ignore{\frac{1}{H}} \int_{\Sigma} d^{2}x\, \Bigg[ &-f_0\, A_{i}\, \partial_{i}\chi + \epsilon_{ij}\, \partial_{t}A_i\, \partial_{j}\chi \\
        &+ v^2 \left( \frac{1}{f_0}A_i - \frac{t}{f_0}\partial_{t}A_i + t\, \epsilon_{ij}A_j \right) \partial_{i} \nabla^2 \chi \Bigg]\, .
    \end{split}
\end{equation}
After performing integration by parts and assuming that all boundary terms vanish, it can be rewritten as follows
\begin{equation}\label{charge-ST-2}
    \inbox{Q_{\chi} = \int_{\Sigma} d^{2}x\, p(t,\vb{x})\, \chi(\vb{x})\, ,}
\end{equation}
where the charge aspect $p(t,\vb{x})$ is given by \footnote{
The charge aspect $p(t,\vb{x})$ depends on the field $A_i$, which might raise concerns about its physical relevance. However, as discussed below \eqref{delta-lambda-S}, $A_i$ cannot be interpreted as a gauge field and does not introduce fake degrees of freedom. Its appearance in \eqref{def-p} is therefore fully consistent.}
\begin{equation}\label{def-p}
    	\ignore{H} p(t,\vb{x}) := f_0\, \nabla \cdot A + \partial_{t}{\eta} - \frac{v^2}{f_0} \nabla^{2} \left( \nabla \cdot A - t  \partial_{t} \nabla \cdot A + t f_0 \eta \right).
\end{equation}
Since $\chi(\vb{x})$ is an arbitrary time-independent function, the conservation of $Q_\chi$ directly implies that $p(t,\vb{x})$ must be \textit{locally} conserved. This can also be verified explicitly by calculating the time derivative of $p(t,\vb{x})$, which yields $\partial_{t} p(t,\vb{x}) = 0$.

Therefore, we have addressed the second question raised in the introduction. In summary, the Noether charge associated with the symmetry transformation Eq. \eqref{symmetry-tranformation} is given by Eq. \eqref{charge-ST-2} with Eq. \eqref{def-p}.

To gain some intuition about the conserved charge Eq. \eqref{charge-ST-2}, we examine its zero, first, and second modes.
\paragraph*{\underline{Zero mode.}} 
It is straightforward to observe that the zero mode of the charge Eq. \eqref{charge-ST-2}, obtained by setting $\chi = 1$, vanishes: $Q_{\chi=1} = 0$.
\paragraph*{\underline{First mode.}} 
Now, let us choose $\chi = \epsilon_{ij} a^{i} x^j$ for an arbitrary constant vector $a^{i}$, so that the charge becomes
\begin{equation}\label{p_i-def}
     Q_{\chi} = a^{i}\int_{\Sigma} d^{2}x\, p_i\, , \qquad 	\ignore{H} p_i := \partial_{t}A_i - f_0\, \epsilon_{ij} A_j\, .
\end{equation}
This charge equals $Q_{\lambda}$ \eqref{Q-lam} for $\lambda=a_{i}x^i$ as one may expect from transformations given by Eq. \eqref{ST-1} and Eq. \eqref{ST-2-M}.

\paragraph*{\underline{Second mode.}} 
To consider the second mode, we take
$$\chi=\frac{-b_{ij}}{4}(\epsilon_{ik}x_{k}x_{j}-\epsilon_{jk}x_{k}x_{i}+\epsilon_{ij}x_k x_k)\, ,$$
where $b_{ij}$ is an arbitrary constant tensor, then 
\begin{equation}\label{Q_ij-def}
     Q_{\chi}=b^{ij}\int_{\Sigma} d^{2}x\, (x_i\, p_j-x_j\, p_i)\, .
\end{equation}
By choosing an appropriate form of $b_{ij}$, we find that the quantity $\frac{d}{dt} \int d^2x\, x_i\, p_j = 0$ is also conserved. 
\paragraph*{Charge algebra.} 
The charge algebra is given by 
\begin{equation}
    \{Q_{\chi_1}, Q_{\chi_2}\}=\delta_{\chi_2}Q_{\chi_1}\, .
\end{equation}
To calculate the charge algebra, we note that $\delta_{\chi} p = 0$. Consequently, the charge aspect $p$ satisfies an abelian algebra as follows
\begin{equation}
            \inbox{
            \{p(\vb{x},t),p(\vb{x}',t)\}=0\, .
            }
\end{equation}
\paragraph*{Non-locality.}
We conclude this section with a brief discussion of \textit{ locality vs. non-locality} of the charge expression Eq. \eqref{charge-ST-2}. One of the central motivations of this work was to identify the conserved charges associated with the symmetry transformation in Eq. \eqref{symmetry-tranformation}. Using the action formulation of the linearized shallow-water symmetry in Eq. \eqref{action-f}, we addressed this question and derived the corresponding conserved charges, as given in Eq. \eqref{charge-ST-2}. A key observation is that obtaining a local expression for these charges requires introducing the symmetry generator $\chi(\vb{x}) $. However, if we instead express the charges in terms of the generator $c(\vb{x})$ using the relation in Eq. \eqref{c-chi-rel}, the resulting charge expression becomes inherently non-local.
\subsection{Spacetime symmetry}\label{sec:spacetime-sym}
In this subsection, we analyze the spacetime symmetries of the shallow water action Eq. \eqref{action-f} with a constant Coriolis parameter $f_{0}$. Specifically, we consider time translation, spatial translation, and rotation.

\paragraph*{\underline{Spatial translation.}}
Under a spatial translation $x^i \to x'^i = x^i + a^i$, where $a^i$ is a constant vector, the shallow water action Eq. \eqref{action-f} transforms as  
\begin{equation}
    \delta_{a} S = \int dt d^{2}x\, a^{i} \partial_{i} \mathcal{L} \, ,
\end{equation}
where we have taken $A_i$ as a vector, with the transformation rule $\delta_{a} A_i = a^{j} \partial_{j} A_i$. Here, $\mathcal{L}$ denotes the Lagrangian density of Eq. \eqref{action-f}.  

The corresponding Noether current is computed as  
\begin{subequations}
    \begin{align}
        &\rho_{a} = \ignore{\frac{1}{H}} 
 a^{k} \bigg[ \partial_{t}A_i\, \partial_{k} A_i +\frac{f_0}{2}\epsilon^{ij} A_i \partial_{k} A_j \bigg] = a^{j} \pi_i \partial_{j} A_i\, ,\\
        &J^{i}_{a} =\ignore{\frac{1}{H}}  a^{k} \bigg[ -v^2 \epsilon^{ij} \eta \partial_{k} A_j + \delta_{k}^{i} \mathcal{L} \bigg]\, .
    \end{align}
\end{subequations}
Thus, the total canonical momentum, as the corresponding Noether charge, is given by  
\begin{equation}
    P_i = \int_{\Sigma} d^{2}x\, \pi_j \partial_{i} A_j\, .
\end{equation}

\paragraph*{\underline{Time translation.}}
Next, we consider time translation. Under a time shift $t \to t' = t + t_0$, the action transforms as  
\begin{equation}
    \delta_{t_0} S = \int dt d^2 x\, t_0 \partial_{t} \mathcal{L}\, .
\end{equation}
The corresponding Noether charge, representing the total energy, is given by  
\begin{equation}
    \mathrm{E} = \int_{\Sigma} d^2 x\, \mathcal{H}\, ,
\end{equation}
where $\mathcal{H}$ denotes the Hamiltonian density Eq. \eqref{Hamiltonian-density}.

\paragraph*{\underline{Rotation.}}
Under an infinitesimal rotation, $x^{i} \to x'^{i} = x^i + \omega^{ij} x^j$, where $\omega_{ij} = -\omega_{ji}$, the shallow water action transforms as  
\begin{equation}
    \delta_{\omega} S = \int dt d^{2}x\, \omega^{ij} \partial_{i}(x_j \mathcal{L})\, .
\end{equation}
The corresponding Noether charge is given by  
\begin{equation}
    J_{ij} = \int_{\Sigma} d^{2}x\, \pi_k (x_i \partial_{j} - x_j \partial_{i}) A_k\, .
\end{equation}
\section{{Applications}}\label{sec:applications}
In this section, we examine the applications and consequences of generalized symmetries in linearized shallow water systems. Specifically, we focus on the subsystem global symmetry {II} given by Eq. \eqref{symmetry-tranformation}.

\subsection{Transformation laws}\label{sec:transformation-laws}
In this subsection, we examine the transformation laws of various quantities under symmetry transformation Eq. \eqref{symmetry-tranformation}
\begin{subequations}
    \begin{align}
        & \Delta_{c}(\nabla \cdot u) = 0\, , \\
        & \Delta_{c} \mathcal{Q}_0 = -	\ignore{H} (\nabla^2 c - R^{-2}c)\, , \label{deltacQ}\\
        & \Delta_{c} \mathcal{H} = -\partial_i (	\ignore{H}\, c \,\epsilon^{ij}\, u_j) + \mathcal{Q}_0\, c + \ignore{H}\frac{1}{2} \left[ (\nabla c)^2 + \frac{c^2}{R^2} \right]\, , \label{deltacE}
    \end{align}
\end{subequations}
where $\Delta_{c} X $ represents the difference in the quantity $X $ before and after the transformation and $R:=\frac{v}{f_0}$ is the Rossby radius of deformation. The first equation shows that the divergence of the velocity field, $\nabla \cdot u $, remains unchanged under the transformation. The transformation law of potential vorticity will play a crucial role in the upcoming sections, particularly in generating solutions via subsystem symmetry.   
\subsection{Subsystem symmetry as a solution generating technique}
One can use the global subsystem symmetry {II} Eq. \eqref{symmetry-tranformation} as a solution-generating technique.  While symmetric transformations generally serve this purpose, subsystem symmetric transformations are particularly notable for their dependence on arbitrary functions. This added flexibility makes them especially powerful and practical. To illustrate their usefulness, we present two examples below.
 
\paragraph*{Geostrophic balance.}  
As a simple example, consider the linear shallow water Eq. \eqref{lin-shallow-eom}, which admits the trivial solution $\eta = 0 $ and $u = 0 $. Using this as a seed solution, we apply the subsystem symmetry {II} Eq. \eqref{symmetry-tranformation} to generate a new class of solutions:  
\begin{equation}
    \eta'(\vb{x}) = -\frac{f_0}{g} c(\vb{x})\, , \hspace{1 cm} u_{i}'(\vb{x}) = \epsilon_{ij} \partial_{j} c(\vb{x})\, .
\end{equation}  
It is straightforward to verify that these are the most general static solutions of the linear shallow water Eq. \eqref{lin-shallow-eom}. The second equation identifies $c(\vb{x}) $ as a stream function for these time-independent solutions. Eliminating $c(\vb{x}) $ allows us to express $\eta' $ and $u_i' $ directly in terms of each other:  
\begin{equation}\label{geostrophic-balance}
    u_{i}'(\vb{x}) = -\frac{g}{f_0} \epsilon_{ij} \partial_{j} \eta'(\vb{x})\, .
\end{equation}  
Steady-state solutions of this form are known as being in \textit{geostrophic balance} \cite{vallis_2017, Fluid-Mech-Book-1, Tong-lectures}.  {It describes a state where the Coriolis force balances the horizontal pressure gradient force in a rotating fluid system.}

 Thus, the symmetry transformation Eq. \eqref{symmetry-tranformation} maps one geostrophic balance solution into another. In other words, these solutions are classified by the conserved charges associated with this symmetry. This highlights that the existence of geostrophic balance solutions is deeply rooted in the underlying subsystem symmetry {II} Eq. \eqref{symmetry-tranformation}.

\paragraph*{Geostrophic Adjustment.}
Geostrophic adjustment is the process through which an initially unbalanced fluid system evolves toward geostrophic equilibrium, where the Coriolis force and the pressure gradient force are in balance. This adjustment is triggered by an initial disturbance—such as a sudden change in surface height or velocity in shallow water models—that disrupts the geostrophic balance. In response, the system generates inertia-gravity waves that propagate outward, carrying away excess energy while leaving behind a slowly evolving, balanced geostrophic flow. Once these transient waves dissipate or exit the region, the fluid settles in a steady-state configuration that satisfies the geostrophic relation $f_0\, {u} = -g\, \hat{z} \times \nabla \eta $. This mechanism plays a crucial role in large-scale atmospheric and oceanic dynamics, influencing phenomena such as Rossby waves, jet stream formation, and oceanic currents, and is extensively studied using shallow water equations and barotropic fluid models \cite{vallis_2017, Fluid-Mech-Book-1, Tong-lectures}.

Suppose we start with an initial configuration given by  
\begin{equation}
    \eta(t=0, \vb{x}) = \eta_0(\vb{x})\, , \qquad u^i(t=0, \vb{x}) = u_0^i(\vb{x})\, .
\end{equation}
The initial data evolves over time according to the dynamical equations, ultimately reaching a new stationary state. Our question is: What is the final configuration of the system?

To address this, we turn to the concept of subsystem symmetries. First, we observe that the two stationary solutions—the initial and final configurations—are related by a global subsystem transformation, as described by Eq. \eqref{symmetry-tranformation}. To determine the final solution, we need only to find the symmetry parameter of the subsystem $c(\mathbf{x}) $. This can be achieved by leveraging the fact that the potential vorticity remains unchanged under transformation, i.e., $\Delta_{c} \mathcal{Q}_0 = 0 $. Consequently, from Eq. \eqref{deltacQ}, we obtain an equation for $c(\mathbf{x}) $
\begin{equation}\label{equation-c}
    (\nabla^2 - R^{-2}) c(\vb{x}) = 0\, .
\end{equation}  
To determine the final stationary state, we first solve this equation for $c(\vb{x}) $ and then apply the subsystem symmetry transformation Eq. \eqref{symmetry-tranformation} to modify the initial configuration accordingly.

Let us consider a concrete example. Suppose the initial condition is given by  
\begin{equation}\label{init-cond-1}
    \eta_{0}(x,y) = h\, \text{sgn}(x)\, , \qquad  u_{0}^i(\vb{x}) = 0\, ,
\end{equation}  
where $h $ is a constant, and the sign function is defined as $\text{sgn}(x>0) = 1 $ and $\text{sgn}(x<0) = -1 $.  The potential vorticity for this configuration is $\mathcal{Q}_0=-f_0\, h\, \text{sgn}(x)$. We further assume that the fluid extends infinitely in the $y $ direction, which implies translational symmetry along $y $, so that all quantities are independent of $y $. Under this assumption, the equation for $c(\vb{x}) $ simplifies to  
\begin{equation}
    \frac{d^2 c}{dx^2} - R^{-2} c = 0\, .
\end{equation}  
The general continuous solution to this equation is  
\begin{equation}\label{solution-c}
    c(x) = c_+ e^{x/R} + c_- e^{-x/R}\, ,
\end{equation}  
where $c_{\pm} $ are constants determined by boundary conditions. The following asymptotic boundary conditions for $\eta$ and $u_i$ 
\begin{equation}\label{boundary-conditions}
    \begin{split}
        &\eta(t = \infty, x = \pm\infty) = h\, \text{sgn}(\pm\infty)\, , \\
        &u_i(t = \infty, x = \pm\infty) = 0\, ,
    \end{split}
\end{equation}  
translate to the following boundary conditions on $c(x)$ as follows
\begin{equation}\label{c-bdry-cond}
    c(x = \pm\infty)=0\, .
\end{equation}
There is no non-trivial continuous solution for $c(x) $ that satisfies the given boundary conditions. This outcome is expected because the initial configuration of the height function contains a discontinuity, while we seek solutions in which the system evolves toward a continuous configuration at the final time. Consequently, the symmetry parameter $c(x) $, which interpolates between these initial and final states, must inherit a discontinuity that matches that of the initial height function. Therefore, $c(x) $ must exhibit the following discontinuity:  
\begin{equation}  
    \text{Discontinuity in } c = \frac{gh}{f_0} \left(1 - (-1)\right) = \frac{2gh}{f_0} \, .  
\end{equation}
Then we obtain  
\begin{equation}\label{GA-c}
    c(x) = \frac{gh}{f_0} \text{sgn}(x) e^{-|x|/R}.
\end{equation}  
Applying the symmetry transformation Eq. \eqref{symmetry-tranformation}, the final state takes the form  
\begin{subequations}\label{final-stationary}
    \begin{align}
        &\eta(t=\infty, x, y) = h\, \text{sgn}(x) (1 - e^{-|x|/R})\, , \label{GA-height}\\
        &u_x(t=\infty, x, y) = 0, \quad u_y(t=\infty, x, y) = \frac{gh}{f_0 R} e^{-|x|/R}\, .
    \end{align}
\end{subequations}  
This example highlights two interesting features. First, although the initial height field configuration exhibits a discontinuity in Eq. \eqref{boundary-conditions}, the system evolves in such a way that the discontinuity vanishes, ultimately reaching a smooth final state. During this process, a portion of the energy propagates to infinity. Second, the parameter of the subsystem symmetry transformation Eq. \eqref{GA-c} itself has a discontinuity. Such discontinuities in symmetry parameters are a common characteristic of systems with subsystem symmetries \cite{Seiberg:2020bhn, Seiberg:2020wsg}.

\paragraph*{Energetic consideration.}
Now, we return to the physical interpretation of the energy transformation law Eq. \eqref{deltacE} in the context of geostrophic adjustment. In this case, the parameter of the symmetry transformation satisfies the differential Eq. \eqref{equation-c}. Using this relation, the energy transformation law Eq. \eqref{deltacE} simplifies to  
\begin{equation}\label{deltacE-GA}
  \Delta_{c}\mathcal{H}=-\partial_{i}(	\ignore{H}\, c\, \epsilon^{ij}\, u_j)+\mathcal{Q}_0\, c+\ignore{H}\frac{1}{2}\nabla\cdot(c\, \nabla c)\, .
\end{equation}  
Since both $c$ and $\mathcal{Q}_0 $ are time-independent, we can rewrite this equation as  
\begin{equation}
    \Delta_{c}\mathcal{H}=-\partial_{i}(	\ignore{H}\, c\, \epsilon^{ij}\, u_j)+\partial_{t}(t\,\mathcal{Q}_0\, c)+\ignore{H}\frac{1}{2}\nabla\cdot(c\, \nabla c)\, ,
\end{equation}  
which can be equivalently expressed in a conservation form:  
\begin{equation}\label{Delta-E-cons}
   \Delta_{c}\mathcal{H}=\partial_{\mu} j^{\mu}\, , \quad j^{\mu}:=\left(t\, \mathcal{Q}_0\, c,-	\ignore{H}\, c\, \epsilon^{ij}\, u_j+\ignore{H}\frac{1}{2} c\, \nabla^{i}c\right)\, .
\end{equation}  
Next, we compute $\Delta_{c}\mathcal{H} $ for the geostrophic adjustment with the initial and boundary conditions given in Eq. \eqref{init-cond-1} and Eq. \eqref{boundary-conditions}. This yields  
\begin{equation}\label{radiation-c-GA}
    \Delta_{c} \text{E}=- L_y\, g\, R\, h^2\, , \qquad \text{E}:= \int_{\Sigma} d^2 x\, \mathcal{H}\, ,
\end{equation}  
where $L_y $ is the length of the system in the $y $-direction and $\text{E}$ is the total energy. To better appreciate the significance of this result, we now perform an energetic analysis of the geostrophic adjustment. The change in potential and kinetic energy is given by  
\begin{equation}  
    \text{V}_{f} - \text{V}_{i} = -\frac{3}{2} L_y\, g\, R\, h^2\, , \quad \text{K}_{f} - \text{K}_{i} = \frac{1}{2} L_y\, g\, R\, h^2\, ,  
\end{equation}  
where $\text{V} =\frac{1}{2} \int d^{2}x\, g\, \eta^2$ and $\text{K} =\frac{1}{2}  \int d^{2}x\, 	\ignore{H} u^2$ denote the potential and kinetic energies, respectively, with the subscripts $i$ and $f$ referring to the initial and final states. From these expressions, we obtain  
\begin{equation}  
    \text{E}_{f} - \text{E}_i = -L_y\, g\, R\, h^2\, .  
\end{equation}  
This result implies that energy is not conserved during the geostrophic adjustment process; instead, a portion of it is radiated away to infinity. As shown in Eq. \eqref{radiation-c-GA}, this energy transformation can be interpreted as
\begin{equation}  
  \text{Energy Radiated} = -\Delta_{c} \text{E}\, .  
\end{equation}  
For detailed computations, see Appendix \ref{app-EC}.
\paragraph*{UV/IR mixing.}
As a final remark in this subsection, we note that the differential equation governing $c(\vb{x})$, given in Eq. \eqref{equation-c}, admits non-continuous solutions, as shown in Eq. \eqref{GA-c}. In other words, imposing a nontrivial boundary condition at large distances (IR), as in Eq. \eqref{c-bdry-cond}, leads to a discontinuity at small distances (UV), Eq. \eqref{solution-c}. This intriguing feature is reminiscent of the UV / IR mixing \cite{TBA}, which naturally emerges in systems with subsystem symmetries \cite{Seiberg:2020bhn, Seiberg:2020wsg, Gorantla:2021bda}.
\section{Summery and Discussion}\label{sec:Discussion}
In this paper, we introduced two subsystem symmetries in the linearized shallow water system, referred to as subsystem symmetry I and subsystem symmetry II. The former leads to the local conservation of potential vorticity and serves as the symmetry foundation of Kelvin's theorem. The latter plays a crucial role in maintaining the geostrophic balance. We also presented an action for the linearized shallow water equations that incorporates a coordinate-dependent Coriolis parameter and examined these symmetries within the framework of Noether’s theorem.

Subsystem symmetries have mainly been studied in exotic field theories, such as fractonic systems \cite{PhysRevLett.94.040402, PhysRevA.83.042330, Vijay:2016phm} (for reviews, see \cite{Nandkishore:2018sel, Pretko:2020cko, Grosvenor:2021hkn, Gromov:2022cxa, Brauner:2022rvf}), which exhibit distinctive features such as system-size-dependent ground state degeneracy, gapped quasiparticles with restricted mobility, UV/IR mixing, and discontinuous symmetry parameters and field configurations. In this paper, however, we demonstrated that these symmetries also emerge naturally in fluid mechanics, particularly in shallow water systems. This raises an intriguing question: Do the characteristic properties of subsystem symmetries also manifest in shallow water dynamics? We plan to investigate this in future work \cite{TBA}.

Looking ahead, we intend to investigate the practical implications of these symmetries and conservation laws. A significant application is the classification and construction of new solutions to the equations of motion. Specifically, the charges associated with the symmetries can serve as labels to distinguish various solutions of the theory. To facilitate this, a systematic construction of the solution space will be pursued in future work.  

Recent developments have revealed that the concept of memory effects, traditionally associated with gauge theories, naturally emerges in fluid mechanics \cite{Sheikh-Jabbari:2023eba}. This raises the intriguing possibility of studying memory effects within the framework of theories possessing subsystem symmetries. Moreover, the interplay between asymptotic symmetries, memory effects, and soft theorems—collectively known as the IR triangle in gauge theories \cite{Strominger:2017zoo}—opens up a rich and promising avenue for further exploration. We aim to investigate the analogous IR triangle in the context of field theories with subsystem symmetries.  

While this work focused on generalized symmetries in effectively $(1+2)$-dimensional shallow water systems, extending these analyses to more general fluid systems represents an exciting direction. Developing an action principle for such systems, as recently achieved in \cite{Taghiloo:2023mtg, Eling:2023apf}, will be a crucial step in this endeavor.  
\section*{Acknowledgement}
We would like to thank Mohammad Mehdi Sheikh-Jabbari for valuable discussions. VT also extends his gratitude to S. Shao and D. Tong for insightful conversations during the Strings 2025 Abu Dhabi conference. Additionally, VT sincerely thanks the organizers of the Strings 2025 Abu Dhabi conference for hosting such an engaging and enriching event. We are also grateful to the referee for comments that helped us avoid misunderstandings and improve the quality of the draft. This work is based upon research funded by Iran National Science Foundation (INSF) under project No. 4040771. 
\appendix
\section{Energetic consideration of the geostrophic adjustment}\label{app-EC}
In this section, we present the detailed computations related to the energetic considerations of geostrophic adjustment discussed in the main text. Our goal is to evaluate the energy transformation given by  
\begin{equation}
     \Delta_{c} \mathcal{H} = -\partial_i (	\ignore{H}\, c \,\epsilon^{ij}\, u_j) + \mathcal{Q}_0\, c + \ignore{H}\frac{1}{2} \left[ (\nabla c)^2 + \frac{c^2}{R^2} \right]\, .
\end{equation}
We begin by considering the expressions for $c \, \mathcal{Q}_0 $ in different regions:  
\begin{equation}
     \begin{split}
         & x>0: \qquad c\, \mathcal{Q}_0 = -gh^2 e^{-x/R}, \\
         & x<0: \qquad c\, \mathcal{Q}_0 = -gh^2 e^{x/R}.
     \end{split}
\end{equation}
First, we note that the total divergence term vanishes because $u_i=0$:  
\begin{equation}
    \int_{\Sigma} d^{2}x\, \partial_i (	\ignore{H}\, c \,\epsilon^{ij}\, u_j) =0\, .
\end{equation}
Next, we compute the contribution from the $c\, \mathcal{Q}_0 $ term:  
\begin{equation}
     \begin{split}
         \int_{\Sigma} d^{2}x \, c\, \mathcal{Q}_0 &= -L_y g h^2  \left[\int_{-\infty}^0 e^{x/R}dx + \int_{0}^{+\infty} e^{-x/R}dx\right] \\
         &= -2 L_y g R h^2.
     \end{split}
\end{equation}
We then evaluate the remaining term. Noting that $\frac{c^2}{R^2} = (\nabla c)^2 $, we write  
\begin{equation}
  \begin{split}
      &x>0: \qquad   \frac{c^2}{R^2} = \left(\frac{gh}{f_0 R}\right)^2 e^{-2x/R}, \\
      &x<0: \qquad   \frac{c^2}{R^2} = \left(\frac{gh}{f_0 R}\right)^2 e^{2x/R}.
  \end{split}
\end{equation}
Thus, integrating over all space, we obtain  
\begin{equation}
    \begin{split}
       \int_{\Sigma} d^{2}x\, &\ignore{H}\frac{1}{2} \qty( (\nabla c)^2 + \frac{c^2}{R^2} )\\ 	=& \ignore{H} L_y \left(\frac{gh}{f_0 R}\right)^2 \bigg(\int_{-\infty}^0 e^{2x/R}dx 
       + \int_{0}^{+\infty} e^{-2x/R}dx\bigg) \\
       =& \ignore{\ignore{H} \frac{L_y}{R} \left(\frac{gh}{f_0}\right)^2 = }L_y\, g\, R\, h^2.
    \end{split}
\end{equation}
Finally, summing the contributions, we arrive at  
\begin{equation}
      \Delta_{c} \int_{\Sigma} d^{2}x\, \mathcal{H} \ignore{= -2 L_y g R h^2 + L_y g R h^2 }= -L_y g R h^2.
\end{equation}
This confirms the result given in Eq. \eqref{radiation-c-GA}.
\section{Forced linearized shallow water}\label{sec:forced-l-sh-w}
In this section, we consider the forced linearized shallow water equations \cite{Sheikh-Jabbari:2023eba} 
\begin{subequations}\label{g-l-sh-w}
\begin{align}
        &\partial_{t}\eta+	\ignore{H} \nabla\cdot u=0\, ,\\
        &\partial_{t}u_i=f \epsilon_{ij} u^j-g\partial_{i}\eta-\partial_i P\, ,
    \end{align}
\end{subequations}
where $P$ represents an external potential. It is straightforward to verify that these equations remain invariant under the symmetry transformation Eq. \eqref{symmetry-tranformation}.  

One of the key motivations for considering forced linearized shallow water systems is the fact that the initial configurations in the geostrophic adjustment problem Eq. \eqref{init-cond-1} do not satisfy the linearized shallow water Eq. \eqref{lin-shallow-eom}. More precisely, while the initial configuration Eq. \eqref{init-cond-1} is a solution to these equations everywhere except at $x = 0 $, a discontinuity arises at this point. The physical reason for this is clear: constructing the initial step-function configuration requires the presence of an external force.  

To account for the role of this external force, we introduce the forced linearized shallow water Eq. \eqref{g-l-sh-w}. Within this framework, the initial configuration becomes a valid solution of the modified equations, with the external forcing term given by $P = -2g h \, \text{sgn}(x) $. This corresponds to a highly localized force, represented by a Dirac delta function $F = 2g h \, \delta(x) $.
		\bibliographystyle{fullsort.bst}

\bibliography{references}
\end{document}